\documentclass{emulateapj}
\usepackage{graphicx}
\usepackage{amsmath}
\usepackage{natbib}
\newcommand{\rl}{RP04}
\newcommand{\gh}{GH09}
\newcommand{\starg}{Star-G}
\newcommand{\hst}{{\it HST}}
\newcommand{\kms}{ km\,s$^{-1}$}
\newcommand{\masyr}{mas\,yr$^{-1}$}

\shortauthors{W.E. Kerzendorf et al.}
\shorttitle{Subaru HDS observations of Tycho Star G}
\begin{document}

\title{Subaru high-resolution spectroscopy of Star G in the Tycho supernova remnant\footnotemark[1]}
\footnotetext[1]{Based in part on data collected at Subaru Telescope,
which is operated by the National Astronomical Observatory of Japan.}
\author{Wolfgang~E.~Kerzendorf} 

\affil{Research School of Astronomy and
Astrophysics, Mount Stromlo Observatory, Cotter Road, Weston Creek,
ACT 2611, Australia}
\email{wkerzend@mso.anu.edu.au}

\author{Brian~P.~Schmidt}
\affil{Research School of Astronomy and
Astrophysics, Mount Stromlo Observatory, Cotter Road, Weston Creek,
ACT 2611, Australia} 
\email{brian@mso.anu.edu.au}

\author{M.~Asplund}
\affil{Max-Planck-Institut f\"{u}r Astrophysik, Karl-Schwarzschild-Stra\ss e 1, Postfach 1317, D-85748 Garching, Germany}

\author{Ken'ichi~Nomoto\altaffilmark{2}}

\affil{Institute for the Physics and Mathematics of the Universe,
 University of Tokyo, 5-1-5 Kashiwanoha, Kashiwa, Chiba 277-8568,
 Japan} 
 \altaffiltext{2}{Department of Astronomy \& Research Center for the Early Universe,
   School of Science, University of Tokyo, Bunkyo-ku, Tokyo 113-0033,
   Japan} 
\email{nomoto@astron.s.u-tokyo.ac.jp}

\author{Ph.~Podsiadlowski} 
\affil{Department of
Astrophysics, University of Oxford, Oxford, OX1 3RH, United
Kingdom} 
\email{podsi@astro.ox.ac.uk}

\author{Anna~Frebel}
\affil{McDonald Observatory, The University of Texas, 1 University Station C1402, Austin, Texas 78712-0259, USA} 
\email{anna@astro.as.utexas.edu}

\and
\author{Robert~A.~Fesen}
\affil{Department of Physics and Astronomy, 6127 Wilder Lab, Dartmouth College, Hanover NH 03755 USA}
\email{fesen@snr.dartmouth.edu}

\author{David Yong} 

\affil{Research School of Astronomy and
Astrophysics, Mount Stromlo Observatory, Cotter Road, Weston Creek,
ACT 2611, Australia}
\email{yong@mso.anu.edu.au}
\begin{abstract}
It is widely believed that Type Ia supernovae (SN Ia) originate in
binary systems where a white dwarf accretes material from a companion
star until its mass approaches the Chandrasekhar mass and carbon is
ignited in the white dwarf's core. This scenario predicts that the
donor star should survive the supernova explosion, providing
an opportunity to understand the progenitors of Type Ia supernovae.In this paper we argue that rotationis a generic signature expected of most non-giant donor stars that is easily measurable. \citep{2004Natur.431.1069R} examined stars in the
center of the remnant of SN 1572 (Tycho's SN) and showed evidence
that a subgiant star (\starg\ by their naming convention) near the
remnant's centre was the system's donor star. We present
high-resolution ($\rm{R} \simeq 40000$) spectra taken with the High Dispersion
Spectrograph on Subaru of this candidate donor star and measure the star's radial velocity as $79\pm 2$\,\kms\ with respect to
the LSR and put an upper limit on the star's rotation of 7.5\,\kms.  In
addition, by comparing images that were taken in 1970 and 2004, we
measure the proper motion of \starg\ to be $\mu_l = -1.6 \pm 2.1$\,\masyr\ and $\mu_b = -2.7 \pm 1.6$\,\masyr. 
We demonstrate that all of the measured properties of \starg\ presented in this paper are consistent with those of
a star in the direction of Tycho's SN that is not associated with the
supernova event. However, we discuss an unlikely, but still viable scenario for \starg\ to be the donor star, and suggest further observations that might be able to confirm or refute it.
\end{abstract}


\section{Introduction}
Type Ia supernovae (SNe Ia) are of broad interest. They serve as
physically interesting end points of stellar evolution, are major
contributors to galactic chemical evolution, and serve as one of
astronomy's most powerful cosmological tools.

It is therefore unfortunate that the identity of the progenitors
of SNe Ia is still uncertain. For example, without knowing the progenitors,
the time scales of SNe Ia enriching the interstellar medium with iron
remains highly uncertain. But it is the crippling impact on the
cosmological application of these objects which is especially
profound; it is impossible to predict the consequences of any
cosmological evolution of these objects or even gauge the likelihood of
such evolution occurring.

There is broad agreement that the stars which explode as SNe Ia are
white dwarfs which have accreted material in a binary system until
they are near the Chandrasekhar mass, then start to ignite carbon
explosively, which leads to a thermonuclear detonation/deflagration of the
star. It is the identity of the binary companion that is currently
completely undetermined. Suggestions fall into two general categories
\citep{1997thsu.conf..111I}:
\begin{itemize}
\item Single degenerate systems in which a white dwarf accretes mass
from a non-degenerate companion, where the companion could be a
main-sequence star, a subgiant, a red giant, or possibly even a
subdwarf.
\item Double degenerate systems where two CO white dwarfs merge,
resulting in a single object with a mass above the Chandrasekhar limit.
\end{itemize}

The detection of circumstellar material around SN~2006X
\citep{2007Sci...317..924P} has provided support for the single
degenerate model in this case, although the lack of substantial 
hydrogen in several other SNe Ia \citep{2007ApJ...670.1275L} 
poses more of a challenge to this scenario.

These models also make different predictions for the nature of the system
following the explosion. In the double degenerate case, no stellar
object remains, but for a single white dwarf, the binary
companion remains largely intact.

In the single degenerate case, the expected effect of the SN on the
donor star has been investigated by \citet*{2000ApJS..128..615M}, who
have calculated the impact of a SN Ia explosion on a variety of binary
companions. \citet*{2001ApJ...550L..53C} have explored many of the
observational consequences of the possible scenarios, and
Podsiadlowski (2003) has presented models that follow both the
pre-supernova accretion phase and the post-explosion non-equilibrium
evolution of the companion star that has been strongly perturbed by
the impact of the supernova shell.  To summarize these results,
main-sequence and subgiant companions lose 10\,--\,20 \% of their
envelopes and have a resulting space velocity of 180\,--\,320\,\kms . Red-giant companions lose most of its hydrogen envelope, leaving a
helium core with a small amount of hydrogen-rich envelope material behind,
and acquire a space velocity of about 10\,--\,100\,\kms.
\citet{2008A&A...489..943P} have used a binary stellar evolution code on a main-sequence star and exposed the evolved star to a SN Ia. Their simulations show that even less material is stripped due to the compact nature of a star that evolved in a binary. We will use their results where applicable. 

\citet[henceforth \rl]{2004Natur.431.1069R} have identified what might
be the donor star to Tycho's SN, a SN Ia which exploded in the Milky
Way in 1572. These authors presented evidence that this star, \starg\
by their naming convention, is at a distance consistent with the Tycho
supernova remnant (henceforth SNR), has a significant peculiar radial
velocity and proper motion, roughly solar abundance, and a surface
gravity lower than a main-sequence star. However, \starg\ is located at a significant distance from the
inferred center of the remnant, and any process that has displaced the
star must preserve the remnant's nearly perfectly circular projected shape. During the final stages of refereeing of this paper we were made aware of the article by \citet[henceforth \gh]{2009ApJ...691....1H}, who used Keck HIRES data to better constrain \starg's stellar parameters, and in addition, found an enhancement in Nickel abundance, relative to normal metal rich stars.

\citet{2007PASJ...59..811I} have looked for Fe absorption lines from
the remnant, using nearby stars as continuum sources, with the hope to
better constrain the distance of these stars to the SNR. With their technique, stars in the
remnant's center should show strong blue-shifted Fe absorption lines,
formed by material in the expanding shell of Fe-rich material from the
SN, moving towards the observer.  Stars in the foreground would show
no Fe absorption, and background stars both red- and blue-shifted
absorption. Their study shows that \starg\ does not contain any
significant blue-shifted Fe absorption lines, suggesting that \starg\
is in the remnant's foreground. However, these observations and their
analysis, while suggestive, cannot be considered a conclusive rebuttal
of \starg's association with the remnant; this technique requires a significant column depth of Fe which is not guaranteed. A lack of Fe column depth may be indicated by the fact that no stars were found in the
vicinity of the remnant that showed both blue- and red-shifted absorption lines.

To further examine the \rl\ suggested association of \starg\ with the
SN Ia progenitor, we have obtained a high-resolution spectrum of the
star using Subaru and its High Dispersion Spectrograph
\citep{1998SPIE.3355..354N}.

We summarize, in section 2, the observational
circumstances of the Tycho remnant and any donor star, and argue in
section 3 that rapid rotation is an important, previously unrealised
signature in a SN Ia donor star. In section 4 we describe our
Subaru observations. Section 5 covers the analysis of data and the results of this analysis. Section 6 compares the relative merit for \starg\ being the donor star to the
Tycho SN or being an unrelated background star, and in section 7 we summarize our findings and motivate future observations.

\section{Observational Characteristics of the Tycho Remnant and Star-G}

\begin{figure}[h!]
\includegraphics*[scale=0.5,angle=-90,width=\linewidth]{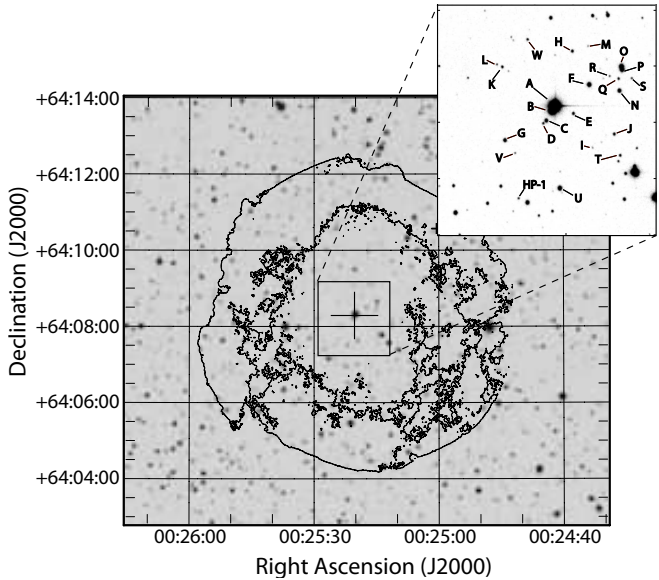}
\caption{Radio Contours (VLA Project AM0347) have been overlaid \citep{1996ASPC..101...80G} on an R-Band Image (NGS-POSS). The cutout is an INT image (see text). The stars marked in the figure are mentioned in this work and in \rl's work. }
\label{fig:overview}
\end{figure}

\rl\ have done a thorough job summarizing the relevant details of the
Tycho remnant. The remnant shows the characteristics expected of a SN
Ia based on its light curve (measured by Tycho Brahe himself),
chemical abundances, and current X-ray and radio emission
\citep{2004ApJ...612..357R}. In figure \ref{fig:overview} we have overlaid radio contours\footnote{The National Radio Astronomy Observatory is a facility of the National Science Foundation operated under cooperative agreement by Associated Universities, Inc.}  on an optical image and have marked the position of the stars mentioned in this and \rl's work.

Although it is not easy to measure the
remnant's distance precisely, \rl\ estimated Tycho's SNR distance to be $2.8 \pm 0.8$\,kpc, using the ratio of the SN 1006 and Tycho SNR's angular sizes and their relative ages, and the direct distance measure of SN 1006 by \citeauthor*{2003ApJ...585..324W} (2003).  \citet{2008Natur.456..617K} have recently shown, from a spectrum of a light echo associated with the SN1572, that this SN was a normal SN Ia. Using Tycho's observed light curve, the properties of SN Ia as standard candles, and an extinction value they find a distance to the SN of $3.8^{+1.5}_{-1.1}$\,kpc. Updating their values for the extinction values determined in this paper (section \ref{sec:distmod}), as well as using an absolute magnitude for SN Ia of $-19.5 \pm 0.25$ \citep{2004MNRAS.349.1344A}, we find a distance of $3.4^{+1.3}_{-1.0}$\,kpc. In summary, we believe the remnant's distance is poorly constrained, but probably between 2 and 4.5\,kpc.
\label{sec:obschar}
\rl\ also report the
spectroscopic and photometric properties for the bright stars near the
center of the Tycho remnant and find a uniform value of approximately
$E(B-V)=0.6$ for stars more distant than 2 kpc. \gh\ have revised the $E(B-V)$ value for \starg\ to 0.76.

In addition, for a select list of stars, \rl\ provide radial velocities and proper
motions. 
For \starg, \rl\ report a value of $v_r=-99\pm6$\,\kms\ for
the radial velocity in the Local Standard of Rest (henceforth LSR), a
proper motion of $\mu_b=-6.1 \pm 1.3$\,\masyr, $\mu_l=-2.6 \pm
1.3$\,\masyr, $\log{g} = 3.5 \pm 0.5$, and $T=5750$\,K.  Using HIRES data \gh\ have improved the measurements of \starg's stellar parameters, finding  $v_r \approx -80$\,\kms, $\log{g} = 3.85 \pm 0.3$, $T=5900 \pm 100$\,K, and $\rm{[Fe/H]}=-0.05 \pm 0.09$\,dex. We note that
\citet{2007PASJ...59..811I} have classified \starg\ as an F8V star ($T
\approx 6250 $\,K, $\log{g} \approx 4.3$,
\citealt{1982lbor.book.....A}), in significant disagreement with the
\rl\ temperature and gravity. We believe the \gh\  values are based on by far the best data, and for the purpose of this paper, we will adopt their values. 

Based on the observations, \rl\ asserted that \starg\ was located at
approximately $3\pm 0.5$\,kpc -- consistent with the remnant's
distance.  They note that this star has solar
metallicity, and therefore its kinematic signature was not
attributable to being a member of the Galactic halo. 
They further argued that \starg's radial velocity and proper
motion were both inconsistent with the distance, a simple Galactic
rotation model, and the star being part of the disk population of the Milky Way.
 The derived physical characteristics of the system were nearly identical to
what was proposed by Podsiadlowski (2003) for a typical SN Ia donor
star emerging from a single degenerate system \citep[e.g., U Sco; also see ][]{Hachisu:1996p758,Li:1997p437,Hachisu:1999p431,Han:2004p444,Han:2008p726}. The revision in the stellar parameters by
\gh\ leads to different distance with a larger uncertainty, but by and large, has not altered the conclusions above. Taken in total, the data provide a rather convincing case for the association of \starg\ with the Tycho SN.

\section{Rapid Rotation: A Key Signature in SN Ia Donor Stars}
\label{sec:rot}
In the single degenerate SN Ia progenitor channel, mass is transferred
at a high rate from a secondary star onto a white dwarf \citep{Nomoto:1982p451,Nomoto:2007p480}. These high
mass-transfer rates require that the secondary star overflows its
Roche lobe. Due to the strong tidal coupling of a Roche-lobe filling
donor, the secondary is expected to be tidally locked to the orbit
(i.e., have the same rotation period as the orbital period).  At the
time of the SN explosion, the donor star is released from its orbit, but
will continue with the same space velocity as its former orbital
velocity and continue to rotate at its tidally induced rate.

\begin{figure}[h!]
\centering
\includegraphics*[width=\linewidth]{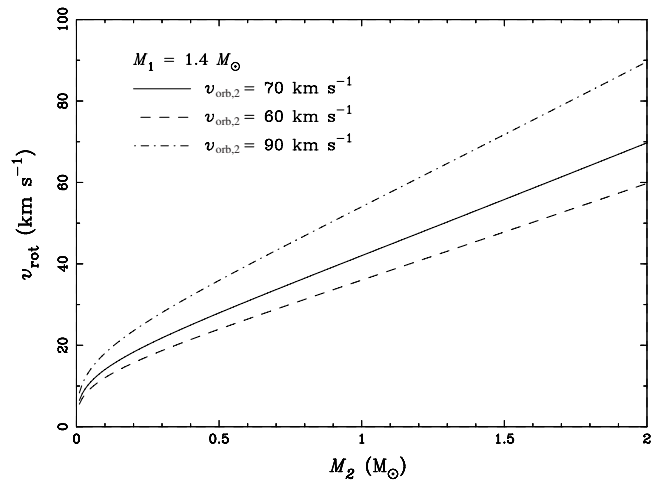}
\caption{The expected rotation rate for a donor star as a function of
its mass at the time of the explosion. The three curves show the results for 3 final space
velocities of the donor star (similar to those suggested by RP04). It
is assumed that the white dwarf has a mass of 1.4\,$M_\odot$.}
\label{fig:theorot}
\end{figure}

There is a simple relationship between the secondary's rotation
velocity $(v_{\rm orb, 2})$ and its orbital velocity:

$$ v_{\rm rot} = {M_1 + M_2\over M_1}\,f(q)\,v_{\rm orb, 2} ,$$

where $f(q)$ is the ratio of the secondary's Roche-lobe radius to the
orbital separation \citep[e.g., given by][]{1983ApJ...268..368E} and
$q=M_1/M_2$ is the mass ratio of the components at the time of the
explosion. Figure \ref{fig:theorot} shows the rotational velocity as a
function of secondary mass for several values of $v_{\rm orb, 2}$ (consistent with \rl s measurement, and at the low end of values expected for a subgiant star),
where we assumed that the exploding white dwarf had a mass of
$1.4\,M_\odot$.

This estimate is strictly speaking an upper limit, as it does not
take into account the angular-momentum loss associated with the
stripping of envelope material by the supernova and any bloating due
to the supernova heating. The latter would reduce the rotational
velocity to first order by a factor equal to the bloating
factor (i.e. the ratio of the new to the old radius), but the
star would likely find itself in a state where its radius and
temperature was atypical of a normal star. 

According to the results of  \citet{2000ApJS..128..615M}, mass stripping is
not likely to be significant if the companion is a main-sequence star
or a subgiant. Furthermore, following binary evolution of a main-sequence star, \citet{2008A&A...489..943P} have shown that even less material is stripped. However, if the companion is a giant, it would be
stripped of most of its envelope.  Such a star would not show any
signs of rapid rotation since the initial giant would have been
relatively slowly rotating; e.g., if one assumes solid-body rotation
in the envelope, the rotation velocity at $\sim 1\, R_\odot$ will only
be $\sim 0.5\,$km\,s$^{-1}$ for a pre-SN orbital period of
100\,d. Moreover, the material at the surface may have expanded from
its original radius inside the giant, further reducing the rotational
velocity. However, if the stripping is less than estimated by  \citet{2000ApJS..128..615M}, then it is possible for the signature of rotation to persist for a giant, albeit at a much lower velocity.

 \citet{2000ApJS..128..615M} also showed that due to the interaction
of the SN blast wave with the companion, the secondary may receive
a moderate kick of up to a few 10\,km\,s$^{-1}$, but this kick
is generally much lower than $v_{\rm orb,2}$ and therefore does
not significantly affect the resulting space velocity.

Finally, we note that the observed rotation velocities are reduced by
a factor $\sin i$, where $i$ is the inclination angle.  However,
because the donor star's rotational axis can be assumed to be parallel
to its orbital axis, a minimum observed rotation speed can be computed
from the observed peculiar radial velocity (observed radial velocity
minus the expected radial velocity of an object at that distance and
direction). It is only if the orbital motion (and hence final systemic
velocity) is solely in the plane of the sky, that $\sin{i}$, and
therefore, the observed rotation, approaches
zero.\label{rotation_expl}

\section{Subaru Observations}
To investigate the rotational properties of \starg, we were granted time
with the Subaru telescope. Our observations of \starg\ were taken in
service mode on the nights of 2005 10 17 and 2005 10 18. 9
spectra were taken with the High Dispersion Spectrograph
\citep[HDS, ][]{1998SPIE.3355..354N} with a resolution of $\rm{R}\simeq40000$ (measured using the instrumental broadening of the Thorium-Argon arc lines), an
exposure time of 2000 seconds each (totalling to 5 hours) and a signal to noise ratio of about 10 per pixel (measured at $8300$\,\AA\ with 0.1 \AA\,pixel$^{-1}$) . The HDS
features two arms, with each arm feeding a 2-chip CCD mosaic. The blue
arm covers 6170\,\AA\ to 7402\,\AA\ and the red arm 7594\,\AA\ to
8818\,\AA. An OG530 filter was used to block contamination from light
blueward of our observing window, and data were binned by 4 in both
the spatial and spectral directions, resulting in a pixel size of
0.1\,\AA\ (at 8000\,\AA) by 0.55$^{\prime\prime}$.

Data were pre-processed using tools provided by the HDS team and then
bias-subtracted. We created a mask from bias and flatfielded frames,
where we isolated the echelle orders and flagged bad pixel
regions. The data were flatfielded using internal quartz flats, and the
2-D images cleaned of cosmic rays (and checked carefully by eye to
ensure there were no unintended consequences) using an algorithm
supplied by M. Ashley (private communication). The spectrum of each
echelle order was extracted using IRAF\footnote{IRAF is distributed by
the National Optical Astronomy Observatory, which is operated by the
Association of Universities for Research in Astronomy (AURA) under
cooperative agreement with the National Science Foundation.}  echelle
routines, with wavelength calibrations based around low-order fits of
a Thorium-Argon arc. Wavelength calibration of each extracted spectrum
was checked against atmospheric O$_2$, and our solutions were found to
be accurate in all cases to within 1\,\kms\
\citep{1985A&A...149..357C}. Unfortunately, we lacked a smooth
spectrum standard star for setting the continuum, and we resorted to
calculating a median of the spectra (6\,\AA\ window) and dividing the
spectra through this smoothed median. This unusual method was chosen
over the common approach of fitting the spectrum with a polynomial,
due to the special characteristics of this observation (low signal to noise ratio, and a complex instrumental response). While this does not affect the narrow lines our program was
targetting, it does affect broad lines such as the H$\alpha$ and the
CaII IR triplet. The final step was to combine all spectra and remove
any remaining cosmic rays (in the 1D spectra) by hand.

\section{Analysis and Results}

\subsection{Rotational measurement}

\begin{figure}[h!]
\centering
\includegraphics*[width=\linewidth]{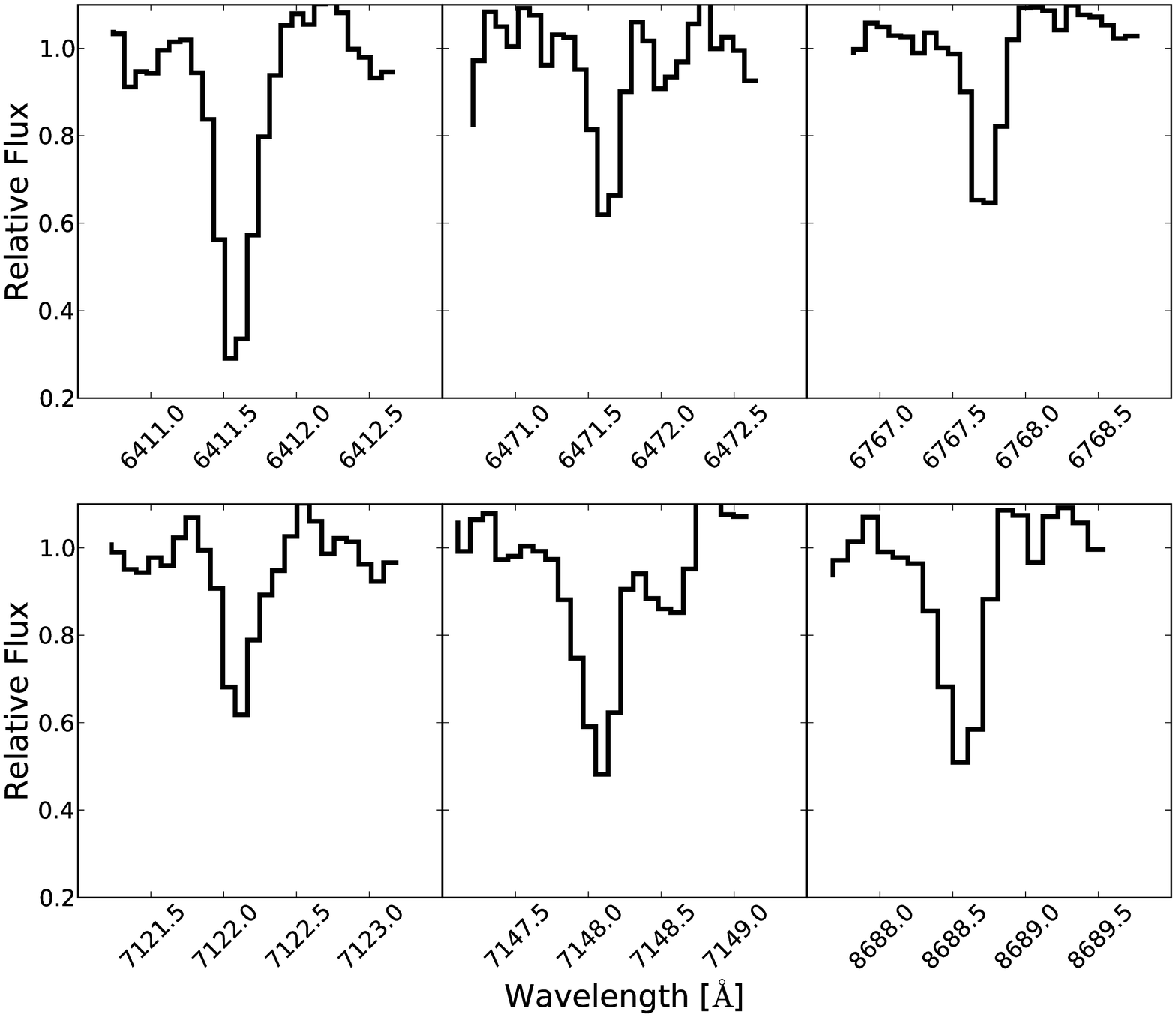}
\includegraphics*[width=\linewidth]{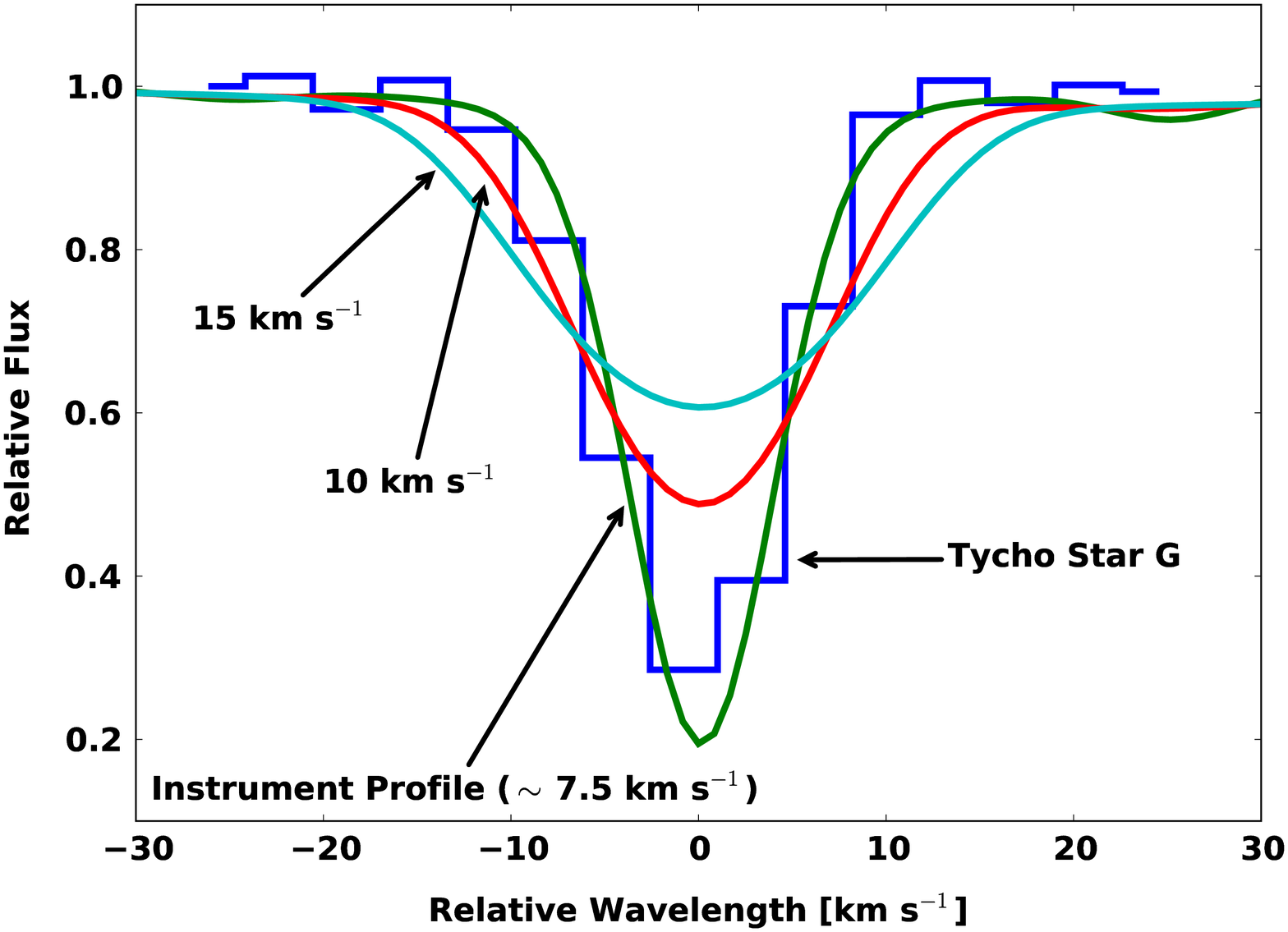}
\caption{Six observed Fe I line profiles of  \starg\ are shown on the left panel. The right panel shows the combination of these line profiles after normalization to the same equivalent width and compares them to the spectrum of the Sun, which is convolved with 3 different values for the rotational broadening kernel. \starg\ does not show significant rotation, indicating $v_{\rm{rot}} \sin{i} \lesssim 7.5$\,\kms.}
\label{fig:sunobjrot}
\end{figure}

To attain the rotational velocity of the candidate star, we measured
several unblended and strong (but not saturated) Fe I lines in the
spectrum \citep{1974lafl.book.....W}. Since our spectrum only had a
combined signal to noise ratio of approximately 10, we added the
spectra of the lines after normalizing them to the same equivalent
width. As a reference we created three synthetic spectra (one broadened only with the instrumental profile, the others with the instrumental profile and $v_{\rm{rot}}\sin{i}$ of 10 and 15\,\kms\ respectively) with the 2007 version of MOOG \citep{1973ApJ...184..839S}, using \gh's temperature, gravity and metallicity.  We use a standard value of $\beta=3/2$ for the limb darkening although the choice
of this value is not critical, which we confirmed by checking our
results using significantly different values of $\beta$. Figure
\ref{fig:sunobjrot} shows the comparison between the synthetic spectra of different rotational velocity and the spectrum of \starg. We have scaled the synthetic spectrum using the equivalent width. This comparison indicates that the stellar broadening (rotational, macro turbulence, etc. ) is less than broadening due to the instrumental profile of 7.5\,\kms, and therefore we adopt 7.5\,\kms\ as our upper limit to the rotation of the star. If one were to adopt \rl's measurements of the
peculiar spatial motion, it could be concluded that $\sin{i}$ is much closer
to 1 than 0 (see the end of section \ref{rotation_expl} for further
explanation) and thus that the rotational speed is $v_{\rm rot} \lesssim 7.5$\,\kms.

\subsection{Radial velocity}

To determine the radial velocity, we used 63 lines to measure the shift
in wavelength. We find a radial velocity in the topocentric (Mauna
Kea) frame of reference of $v_{\rm top} = - 92.7 \pm 0.2$\,\kms (the
error being the standard deviation of 63 measurements) . The
conversion from the topocentric to the Galactic LSR for our
observations was calculated to be 13.6 \kms\ (IRAF task rvcorrect)
using the IAU standard of motion. Including the uncertainty in the LSR
definition, we find a radial velocity in the LSR for \starg\ of
$v_{\rm LSR} = -79 \pm 2$\,\kms. This is in significant disagreement
with that reported by \rl, but agrees with the revised value published by \gh.

\subsection{Astrometry}
\rl\ have measured a significant proper motion for \starg\ of $\mu_b=-6.1 \pm 1.3$ \masyr, $\mu_l=-2.6 \pm 1.3$\ \masyr. Because \starg\ is metal rich, and at a distance of $D>2$kpc, this measurement provides one of the strongest arguments for \starg\ being the donor star to Tycho SN. It is almost impossible to account for this proper motion, equivalent to a $v_b=58\left({{D}\over{2\rm{~kpc}}}\right)$\kms\ or 3 times the disk's velocity dispersion of $\sigma_z=19$\kms, except through some sort of strong binary star interaction.  

However, the \hst\ data present an especially difficult set of issues in obtaining astrometry free of systematic errors. For \starg\ these issues include the PSF on the first epoch WFPC2 image being grossly undersampled, both the ACS and WFPC2 focal planes being highly distorted,  poor and different charge transfer efficiency across the two \hst\ images, and that \starg\ was, unfortunately, located at the edge of one of the WFPC2 chips, making it especially difficult to understand the errors associated with it. Smaller issues include the small field of overlap between the two images, making the measurement subject to issues of the correlated motions of stars, especially in the $\mu_l$ direction.

To cross-check \rl's proper motion of Star-G , we have scanned a photographic plate taken in
September 1970 on the Palomar 5 meter, and compared this to an Isaac
Newton 2.5\,m Telescope (INT) CCD archive image (INT200408090414934) of the remnant taken in
August 2004. The Palomar plate has an image FWHM of $1.7^{\prime\prime}$, and the INT image $0.88^{\prime\prime}$. While our images have a much larger PSF than the HST images, the images have significantly less distortion, are matched over a larger field of view with more stars, have fully sampled PSFs, and were taken across nearly an 8 times longer time baseline. The photographic nature of the first epoch does add complications not present in the \hst\ data. The non-linear response of photographic plates causes their astrometry to have systematic effects as a function of brightness \citep{2001ASPC..232..311C}, especially affecting objects near the plate limit, where  single grains are largely responsible for the detection of an object.

The position of stars on the INT image were matched to
the 2MASS point source catalog \citep{2006AJ....131.1163S} to get a coordinate 
transformation (pixel coordinates to celestial coordinates) using a
3rd-order polynomial fit with an RMS precision of 40 mas with 180 stars. This fit is limited by precision of the 2MASS catalog and shows no systematic residuals as a function of magnitude, or position.  Using this  world coordinate system (WCS) transformation, we then derived the
positions of all stars on the INT image. The coordinates of 60 uncrowded
stars on the Palomar plate were matched to the INT-based catalog, and a 3rd-order polynomial was used to transform the Palomar positions to the INT-based positions. The fit has an RMS of 65 mas in the direction of galactic longitude, and 45mas in the direction of galactic latitude.  We believe the larger scatter in the direction of  Galactic longitude is due to the shape of the PSF being slightly non-symmetric in the direction of tracking on the Palomar plate. This tracking (in RA, which is close to the direction of galactic longitude), causes the position of stars to depend slightly on their brightness. This explanation is supported by  a small systematic trend in our astrometric data in $\mu_l$, not seen in $\mu_b$, as a function of $m_R$.  An alternative explanation is that the  trend in $\mu_l$ is caused by the average motion of stars changing due to galactic rotation as a function of distance, which is proxied by $m_R$.  We have used the Besan\c{c}on Galactic model \citep{2003A&A...409..523R} to estimate the size of any such effect, and find the observed effect is an order of magnitude larger than what is expected. The systemic difference between assuming either source of the observed effect is less than 1\,\masyr\ in $\mu_l$, and has no effect in our $\mu_b$ measurement. In our final proper motions, presented in table \ref{tab:prop_motion}, we remove  the systematic trend as a function of $m_R$ with a linear function.

\begin{table*}[t!]
\caption{Proper motions of stars within $45^{\prime\prime}$ of the Tycho SNR center.}
\begin{center}
\begin{tabular}{ccccccc}
\hline\hline														
$\alpha$ &	$\delta$	&	$\mu_l$	&	$\mu_b$	&	$m_R$	&	$\theta$ \\		
\,[hh:mm:ss.ss] & [dd:mm:ss.ss] & [\masyr] & [\masyr] & [mag] & [arcsec] & Name \\
%RA                                     & DEC                                      & Proper motion in $l$      &      Proper motion in $b$ &   apparent magnitude in R (\rl) & distance from center & designation given by \rl \\
\hline
														
00:25:20.40	&	+64:08:12.32	&	-0.90	&	-0.56	&	17.05	&	08.9	&	c	\\	
00:25:18.29	&	+64:08:16.12	&	-4.25	&	-0.81	&	18.80	&	10.0	&	e	\\	
00:25:17.10	&	+64:08:30.99	&	-1.82	&	1.78	&	16.87	&	20.3	&	f	\\	
00:25:23.58	&	+64:08:02.02	&	-1.58	&	-2.71	&	17.83	&	31.1	&	g	\\	
00:25:15.52	&	+64:08:35.44	&	1.94	&	0.83	&	20.28	&	31.4	&	r	\\	
00:25:15.08	&	+64:08:05.95	&	-0.67	&	1.49	&	18.86	&	33.3	&	j	\\	
00:25:23.89	&	+64:08:39.33	&	-0.31	&	1.08	&	19.20	&	33.5	&	k	\\	
00:25:14.74	&	+64:08:28.16	&	2.60	&	1.46	&	17.45	&	33.5	&	n	\\	
00:25:14.81	&	+64:08:34.22	&	4.05	&	-2.05	&	19.35	&	35.0	&	q	\\	
00:25:13.79	&	+64:08:34.50	&	2.32	&	1.01	&	19.90	&	41.3	&	s	\\	
00:25:14.59	&	+64:07:55.10	&	-3.94	&	2.35	&	19.23	&	41.7	&	t	\\	
00:25:19.25	&	+64:07:38.00	&	1.75	&	-3.43	&	16.86	&	42.1	&	u	\\	
00:25:22.45	&	+64:07:32.49	&	81.29	&	-2.68	&	19.81	&	48.7	&	HP-1	\\	\hline
\end{tabular}
\end{center}
\label{tab:prop_motion}

\end{table*}

To measure the proper motion of each star, we exclude each star from the astrometric transformation fit
so as not to bias its proper motion measurement.  Comparing the stellar positions in the 34 year interval we find that these 60 stars show an RMS dispersion  $\sigma_{\mu_l} = 2.1$\masyr, $\sigma_{\mu_b} =1.6$\,\masyr. For
\starg\ we measure $\mu_l = -1.6 \pm 2.1$\,\masyr, $\mu_b = -2.7 \pm 1.6$\,\masyr; this implies that no significant proper motion is detected. We do note that this measurement has a similar precision to
that of \rl, is consistent with no observed motion, and is in moderate disagreement with the \rl\ measurement.

\begin{figure}[h!]
	\centering
	\includegraphics[width=\linewidth]{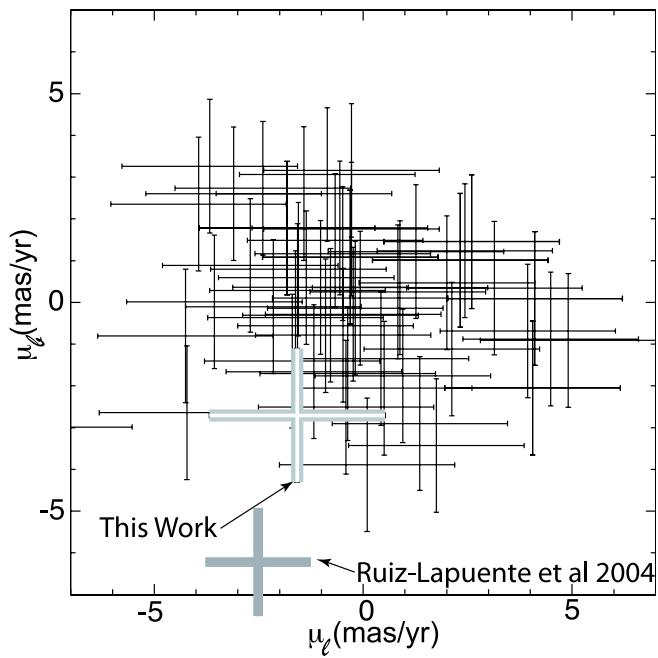}
\caption{The astrometric motions of 60 stars measured in the Tycho SNR
center. The measurements have a RMS dispersion of 1.6\,\masyr. Shown
in grey is the proper motion of \starg\ measured here and by RP04, showing a moderate discrepancy in the two measurements. Our measurement is consistent with no proper motion. }
\label{fig:prop_motion}
\end{figure}

In table \ref{tab:prop_motion} we present our astrometric measurements of all stars listed by \rl\  for which we were able to measure proper motions. We also give the apparent magnitudes in R (partly measured by this work and partly by \rl) and the distance from center $\theta$. Due to crowding caused by the relatively poor resolution of the first epoch photographic plate, several stars are not included that could be measured using \hst.
We include an  additional star, not cataloged by \rl, which exhibits high proper motion. This  high proper motion star, which was off the WFPC2 images of \rl,  we designate HP-1, and has a proper motion of $\mu_l=81.3$, $\mu_b=-2.7$ \masyr. Due to the distance from the remnant's center, (we estimate HP-1 would have been located 51$^{\prime\prime}$ from the remnant's center in 1572), we doubt this star is connected to the Tycho SN, but we include it for the sake of completeness. 

\section{Discussion}

\subsection{A Background interloper?}
A previously unrecognized property for many progenitor scenarios is the rapid post-explosion rotation of the donor (as described in section \ref{sec:rot}).
The expected rotation as calculated in Figure \ref{fig:theorot} is
large compared to that expected of stars with a spectral type later
than F and should be easily observable. We have shown \starg's rotation to be less ($v_{\rm rot} \sin{i} \lesssim 7.5 $\,\kms) than what is expected  of an associated star if the companion was a main-sequence star or subgiant. A red giant scenario where the envelope's bloating has significantly decreased rotation could be consistent with our observation of \starg, and this will be discussed in section \ref{redgiant}.

\begin{figure}[htb!]
\centering
\includegraphics* [scale=0.5,angle=0]{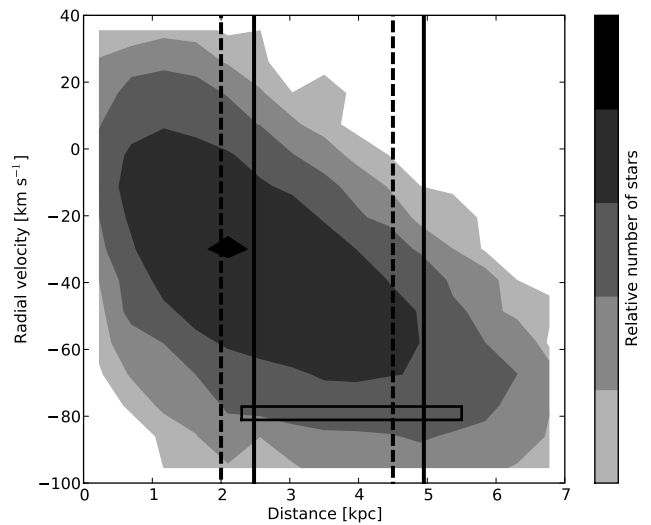}
\caption{Besan\c{c}on model for a metal rich ([Fe/H] $>$ -0.2 ) Galactic population between 0 and 7\,kpc in the direction of Tycho SNR (l = 120.1, b=1.4) with a solid angle of 1 square degree.
The remnant's distance is represented by the black dashed lines (as calculated in section \ref{sec:obschar}). The contours show the radial velocity distribution. 
Our measured radial velocity corrected to LSR and our distance are shown, with their respective error ranges, as the black rectangle.  The distance  range calculated by \gh\ are indicated by the two solid lines. The observed LSR $v_r$ for \starg\ is mildly unusual for stars at the remnant's distance, and is consistent with the bulk of stars behind
the remnant. }
\label{fig:bes_d_mu}
\end{figure}

The primary basis for which \rl\ selected \starg\ as a candidate for
the donor star to the Tycho SN was the combination of its large peculiar radial velocity
and its observed proper motion. In Figure \ref{fig:bes_d_mu} we
use the Besan\c{c}on Galactic model \citep{2003A&A...409..523R} to
construct an expected set of radial velocities for metal-rich stars in the
direction of SN1572.

Measuring the distance to \starg\ is a key discriminant in associating the star to the SN explosion. To improve the uncertainty of the distance to the star, due both to temperature and extinction uncertainty,  we base our distance on the observed $m_K$ \citep{2006AJ....131.1163S} and ($V-K$) color (\rl).  We interpolate ATLAS9 models without overshoot \citep*{1998A&A...333..231B} to find a theoretical $V-K$ and absolute magnitude for the \gh's values of temperature and gravity. Using a standard extinction law \citep*{1989ApJ...345..245C} ($A_V= 3.12 E(B-V)$ and $A_K/A_V=0.109$) to match the theoretical and observed colors, we find $A_V=2.58\pm0.08$mag, $A_K=0.28\pm 0.01$ mag, and $E(B-V)=0.84\pm0.05$.  To better show the uncertainties, we present our distance moduli scaled to the observed and derived values of extinction, temperature and gravity.
The temperature coefficients were determined by integrating blackbodies of the appropriate temperature with a filter bandpass and fitting a powerlaw to the resulting flux. \label{sec:distmod}
 
\begin{equation}
\nonumber
\begin{split}
(m_V-M_V)=&12.93-3.12(E(B-V)-0.84)\\
&-2.5 (\log{g} -3.85)+2.5\log\left(\frac{M}{1\,M_\odot }\right)\\
&+2.5\log\left( \frac{T_{\rm{eff}}}{5900} \right)^{4.688}\\
%\end{split}
%\begin{split}
(m_K-M_K)=&12.93-0.275(E(B-V)-0.84)\\
&-2.5 (\log{g} - 3.85)++2.5\log\left(\frac{M}{1\,M_\odot }\right)\\
&+2.5\log\left( \frac{T_{\rm{eff}}}{5900} \right)^{1.937}
\end{split}
\end{equation}

Assuming a companion mass of $1\,M_\odot$ we find a $(m-M)=12.93\pm0.75$ mag. This uncertainty is dominated by the precision of $\log{g}$, and equates to a distance of $D=3.9\pm1.6$kpc. \starg, within the errors, is at a distance consistent with the remnant.  As seen in Figure \ref{fig:bes_d_mu}, the observed radial velocity of \starg\ is consistent with a significant fraction of  stars in its allowed distance range.  We also note that if \starg\ is indeed associated with the SN, that it is likely that \starg\ could have a mass considerably less than $1\,M_\odot$, due to mass transfer and subsequent interaction with the SN, although in this case, the distance to the star would still be consistent with SNR distance.

\citet{2007PASJ...59..811I} looked for absorption due to Fe I in the remnant's expanding ejecta for 17 stars within the Tycho remnant.  No such absorption was seen in the spectrum of \starg, potentially placing it in front of the remnant. However, the amount of Fe I currently within the remnant is uncertain with predicted column densities spanning several orders of
magnitude \citep[$0.02 - 8.9 \times 10^{15}\rm{\,cm}^{-2}$;][]{Hamilton:1988p522,Ozaki:2006p517}. Therefore, we do not believe the lack of significant Fe I 3720 absorption in \starg\ to be significant.

In summary, we find that \starg's radial velocity, distance, and stellar parameters are all consistent with an unrelated star, but also with it being the donor star. There is disagreement in  \starg's measured proper motion. The measurements of \rl\ are inconsistent with normal disk stars at the known distance and strongly point to \starg\ being associated with the SN, whereas the measurements presented here are consistent with a normal disk star, unrelated to the SN. In addition, we have shown the rotation of \starg\ is low (confirmed by \gh ; $v_{\rm{rot}} \leq 6.6$\,\kms  ), arguing against association with the SN, as does its off center placement in the remnant.  Finally, \gh\ have presented evidence that \starg\ is strongly enhanced in Nickel, an observation that, if confirmed, would strongly point to an association of the star with the SN. If either the high proper motion, or significant Nickel enhancement can be confirmed, then it is likely that \starg\ is the SN donor star. Otherwise, we believe it is much more likely that \starg\ is simply an interloper.

\subsection{\starg\ as the Donor Star to the Tycho SN}
\label{redgiant}

While the case for \starg's association with the SN is not conclusive, it is intriguing, and we believe it is worthwhile to look for a consistent solution assuming the association is true.
While not {apriori} probable, a self-consistent model can be constructed in which \starg\ was the companion, as we shall discuss now.

To make such a model work, \starg\ has to be a stripped giant that
presently mimics a G2IV star. At the time of the explosion, the star
would have been a moderately evolved giant (in a binary with an
orbital period $\sim 100\,$d). The SN ejecta will strip such a giant
of almost all of its envelope \citep{2000ApJS..128..615M} due to its low binding energy; only the most tightly bound envelope material
outside the core will remain bound. Due to the heating by the SN,
even this small amount of material (perhaps a few $\times
0.01\,M_{\odot}$) will expand to giant dimensions, and the
immediate-post-SN companion will have the appearance of a luminous red
giant. However, because of the low envelope mass, the thermal
timescale of the envelope is sufficiently short that it can loose most
of its excess thermal energy in 400 years and now have the appearance
of a G2IV star \citep{2003astro.ph..3660P}.

A lower mass for \starg\ ($0.3-0.5\,M_\odot$) also
reduces the distance estimate, and makes the observed radial velocity more unusual for stars at this distance.
The expected spatial velocity depends on the
pre-SN orbital period and should be in the range of
$30-70\,$km\,s$^{-1}$ for a period range of $20-200\,$d \citep[]{Justham:2008nx}. These velocities are consistent with the
inferred spatial velocity of the object relative to the LSR if \starg\
is at the distance of the remnant, even if no significant proper motion has been measured (see Figure \ref{fig:bes_d_mu}).

A stripped-giant companion would link the progenitor to the symbiotic
single-degenerate channel \citep{1999ApJ...522..487H} for which the
symbiotic binaries TCrB and RS Oph are well studied
candidates. Indeed, \citep[]{Justham:2008nx} argued that
the ultracool low-mass helium white dwarfs (with masses $\la
0.3\,M_\odot$) that have been identified in recent years are most
likely the stripped-giant companions that survived SN Ia explosions, 
which could provide some further possible support for such a scenario for
\starg.

If the association is real, \starg's displacement to the SE of the geometric
center of the remnant as defined by radio and X-ray observations might
be interpreted as being due to the remnant's interaction with an
inhomogeneous ISM.  Deep optical images of the remnant do show
extended diffuse emission along the eastern and northeastern limbs
interpreted as shock precursor emission
\citep{2000ApJ...535..266G}. This along with an absence of detected
Balmer-dominated optical emission along the whole of the western and
southern limbs suggests a density gradient of the local interstellar
medium with increasing density towards the NE. An east-west density
gradient has also been inferred from detailed radio expansion rate
measurements \citep{1997ApJ...491..816R}.  Such an E--W density
gradient could have led to a more rapid expansion toward the west
giving rise to a small shift in the apparent geometric center away
from the SE without creating a highly distorted remnant.  However, there are problems with this explanation. Deviations from spherical symmetry in both radio and X-ray
images of the remnant are relatively small
\citep{1997ApJ...491..816R,2007ApJ...665..315C}, and the remnant is
most extended along the eastern and northeastern limbs, just where one
finds the greatest amount of extended diffuse optical
emission.
Moreover, the remnant's expansion rate appears lowest toward
the northeast (PA = 70 degrees), not the southeast \citep{1997ApJ...491..816R}. Although the
argument that \starg's SE displacement from the remnant's current
geometric center is a result of an asymmetrical expansion is not
strong, it remains a possibility.

The most conclusive way of confirming a stripped-giant scenario for
\starg\ would be an independent, precise measurement of the distance
to \starg\ which in combination with measurements of the gravity and
effective temperature would help to constrain \starg's
mass. Unfortunately, such a measurement will most likely have to wait
for the advent of the GAIA satellite.  Alternatively, one may be able
to single out a stripped giant from a normal G2IV star through
nucleosynthesis signatures, specifically evidence for CNO-processed
material (or other nucleosynthetic anomalies).  While a normal G2IV star is unlikely to show CNO-processed
material at the surface, a stripped giant is likely to do so. Unfortunately, the data presented here are not of adequate quality to explore the detailed properties of \starg's atmosphere.

\section{Outlook and Future Observations}

Presently, we believe the evidence for \starg's association with the Tycho SN is interesting, but not conclusive.
A possible scenario if \starg\ is the donor star, would be that of a stripped giant scenario discussed in section 6.  
However, there are still other stars that have not been adequately
scrutinized. \citet{2007PASJ...59..811I} have found a star (\rl\
Star-E) which may contain blueshifted Fe I lines, indicating their
association with the remnant. Unfortunately, the star has neither
a significant peculiar radial velocity (\citealt{2007PASJ...59..811I}; \rl)
 nor a significant peculiar proper motion (\rl\ and confirmed by
our work; see Table \ref{tab:prop_motion}).

High-resolution spectroscopy of each candidate in the remnant's center is necessary to
precisely determine each star's physical parameters. However, the small observed
velocities of the remaining stars suggest that the donor star would
have needed to be a giant at the time of explosion. Using \rl's
observed values, none of the stars in the remnant's center appear
consistent with what is expected of a giant star as the donor star
except possibly for Star-A.  We also note that there is an additional
star present in archived HST images, not cataloged in \rl, offset from \rl's star A  by
0.5$^{\prime\prime}$ E and 0.2$^{\prime\prime}$ N at $m_V=16.8$,
$(B-V)=1.0$. This star, near the remnant's centre, has a color
consistent with an F-star (assuming that it is behind the bulk of the
line of sight reddening), but it will require adaptive optics to
obtain its spectrum given its proximity to the 13th magnitude Star-A. This star could potentially be a non-giant
progenitor.

If future observations are unable to pinpoint a viable donor star,
other progenitor scenarios will have to be considered. These include
the double degenerate scenario, or a scenario where there is a long
time delay between the accretion phase of a donor star onto the white dwarf, and the ultimate supernova
explosion.

\bigskip
We would like to thank the Subaru HDS team for taking these
observations in service mode. This paper makes use of data obtained
from the Isaac Newton Group Archive which is maintained as part of the
CASU Astronomical Data Centre at the Institute of Astronomy,
Cambridge. This publication makes use of data products from the Two
Micron All Sky Survey, which is a joint project of the University of
Massachusetts and the Infrared Processing and Analysis
Center/California Institute of Technology, funded by the National
Aeronautics and Space Administration and the National Science
Foundation. This work also makes use of POSS-I data. The National Geographic Society - Palomar Observatory Sky Atlas (POSS-I) was made by the California Institute of Technology with grants from the National Geographic Society. 
WEK, BPS and MA are supported by the Australian Research
Council (grant numbers DP0559024, FF0561481). This paper was conceived as part of the Tokyo Think Tank
collaboration, and was supported in part by the National Science
Foundation under Grant No. PHY05-51164.
This work was supported in part by World Premier International
Research Center Initiative (WPI Program), MEXT, Japan, and by the
Grant-in-Aid for Scientific Research of the Japan Society for the
Promotion of Science (18104003, 18540231, 20540226) and MEXT (19047004, 20040004). 
Additionaly we would like to thank Pilar Ruiz Lapuente and her team for the valuable discussions we had in regards to the manuscript. We would also like to thank our referee, who provided us with a very detailed and thorough analysis of the first manuscript and subsequent revisions. 

\bibliographystyle{hapj}

\end{document}